\documentclass{iaus}
\usepackage{graphicx}

\title[MHD Mode Conversion]{MHD mode conversion in a stratified atmosphere}

\author[A.~M.~D.~McDougall \and A.~W.~Hood]{A.~M.~Dee McDougall \and Alan W.~Hood}

\affiliation{School of Mathematics and Statistics, University of St Andrews, St Andrews, KY16 9SS, UK \break email: dee@mcs.st-andrews.ac.uk}

\pubyear{2007}
\volume{247}
\pagerange{296--302}
\date{10.1017 and in revised form ??}
\jname{Waves and Oscillations in the Solar Atmosphere: \\Heating and Magneto-Seismology}
\editors{R.~Erd\'elyi \& C.~A.~Mendoza-Brice\~no, eds.}
\doi{10.1017/S1743921308014993}

\begin{document}

\maketitle

\begin{abstract}
Mode conversion in the region where the sound and Alfv\'en speeds are equal is a complex process, which has been studied both analytically and numerically, and has been seen in observations.  In order to further the understanding of this process we set up a simple, one-dimensional model, and examine wave propagation through this system using a combination of analytical and numerical techniques.  Simulations are carried out in a gravitationally stratified atmosphere with a uniform, vertical magnetic field for both isothermal and non-isothermal cases.  For the non-isothermal case a temperature profile is chosen to mimic the steep temperature gradient encountered at the transition region.  In all simulations, a slow wave is driven on the upper boundary, thus propagating down from low-$\beta$ to high-$\beta$ plasma across the mode-conversion region.  In addition, a detailed analytical study is carried out where we predict the amplitude and phase of the transmitted and converted components of the incident wave as it passes through the mode-conversion region.  A comparison of these analytical predictions with the numerical results shows good agreement, giving us confidence in both techniques.  This knowledge may be used to help determine wave types observed and give insight into which modes may be involved in coronal heating.  
\end{abstract}

\firstsection 

\section{Introduction}
\label{intro.sec}
Mode conversion is the process under which an incident wave may be wholly or partially converted into another wave mode, with the remainder of the original wave being transmitted through the plasma.  This may occur where the sound and Alfv\'en speeds are equal, or equivalently, where the plasma $\beta$ (the ratio of the gas to the magnetic pressure) is approximately unity.  This layer will generally lie low down in the solar atmosphere, in the region of the lower chromosphere.  

This problem has been investigated extensively in the past, many using analytical techniques.  \cite{Zhugzhda1979}, \cite{Zhugzhda1981} and \cite{Zhugzhda1982} investigated the conversion of fast magnetoacoustic waves into slow magnetoacoustic waves as they propagate up through the solar atmosphere from high- to low-$\beta$ plasma.  Their model looked at a uniform vertical magnetic field set in an isothermal atmosphere stratified by gravity.  Using the fact that, for harmonic waves, an exact solution may be found in terms of the known Meijer-G functions, coefficients describing the conversion and transmission were found.  \cite{Cally2001} expanded on this work by noting that these solutions may also be written in terms of the equivalent Hypergeometric $_2F_3$ functions, and another set of coefficients was discovered.  

\begin{figure}[t]
\centering
\includegraphics[scale=0.5]{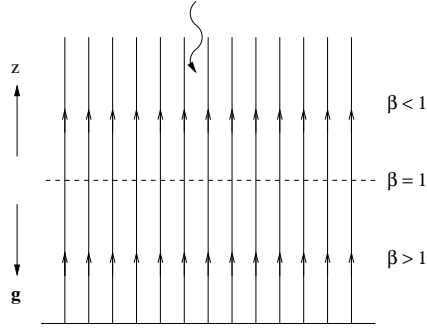}
\caption{Cartoon of our model atmosphere, with uniform vertical magnetic field.  The $z$-axis points upwards (opposite to gravity) and a slow wave is driven on the upper boundary, travelling down towards the mode-conversion layer at $\beta\approx1$.}
\label{model.fig}
\end{figure}

We also investigate the mode conversion problem analytically, although we combine this with numerical simulations.  We look at a uniform background magnetic field, but, in contrast to the aforementioned work, we investigate downward propagation - so the wave is travelling from low- to high-$\beta$ plasma (as shown in Figure~\ref{model.fig}).  As the fast wave is evanescent in low-$\beta$ plasma, we look at the conversion of slow to fast magnetoacoustic waves.  

In Section~\ref{eqns.sec} the basic equations are set out, and we look at the characteristic wave behaviour in high- and low-$\beta$ plasma.  Two different situations are then examined: an isothermal atmosphere in Section~\ref{iso.sec}, and a non-isothermal atmosphere in Section~\ref{noniso.sec}.  Finally the conclusions are detailed in Section~\ref{conc.sec}.  

\section{Basic Equations}
\label{eqns.sec}

Throughout we use the standard, ideal MHD equations, linearised about the equilibrium ${\bf B}_0=B_0{\bf\hat{k}}$, and $p_0^\prime=-\rho_0g$; where $p_0=\rho_0RT_0\left(z\right)$ and $H=g/RT_0$ may be defined as the scale height.  So we have the equation of motion
\begin{equation}
\rho_0\frac{\partial{\bf v}_1}{\partial t}=-{\bf\nabla}p_1+\frac{1}{\mu}\left({\bf\nabla}\times{\bf B}_1\right)\times{\bf B}_0+\rho_1{\bf g},
\label{linmo.eqn}
\end{equation}
the induction equation
\begin{equation}
\frac{\partial{\bf B}_1}{\partial t}={\bf\nabla}\times\left({\bf v}_1\times{\bf B}_0\right),
\label{linind.eqn}
\end{equation}
the mass continuity equation
\begin{equation}
\frac{\partial\rho_1}{\partial t}+{\bf\nabla}\cdot\left(\rho_0{\bf v}_1\right)=0,
\label{lincont.eqn}
\end{equation}
and the energy equation
\begin{equation}
\frac{\partial p_1}{\partial t}+\left({\bf v}_1\cdot{\bf\nabla}\right)p_0=\frac{\gamma p_0}{\rho_0}\left(\frac{\partial\rho_1}{\partial t}+\left({\bf v}_1\cdot{\bf\nabla}\right)\rho_0\right),
\label{linen.eqn}
\end{equation}
where subscript zero denotes the equilibrium quantities and subscript one the perturbed quantities.  In these equations ${\bf B}_0$ is the equilibrium magnetic field, and $p_0$ and $\rho_0$ are the equilibrium pressure and density which may be calculated from the equilibrium equation above.  The variable ${\bf v}_1$ denotes the plasma velocity, $p_1$ is the plasma pressure, $\mu$ is the magnetic permeability, ${\bf B}_1$ is the magnetic field, $\rho_1$ is the mass density, ${\bf g}=-g{\bf\hat{k}}$ is the gravitational acceleration, and $\gamma$ is the ratio of specific heats, which we take to be $5/3$.  Henceforth, the subscripts on the perturbed variables are dropped, and it is understood that we are working with the linearised equations.  

Under the assumption that the $x$-dependence is of the form $\exp{ik_xx}$, where $k_x$ is\break the horizontal wavenumber, Equations~(\ref{linmo.eqn})\,--\,~(\ref{linen.eqn}) may be combined to form a pair of wave equations
\begin{equation}
\left(v_A^2\frac{\partial^2}{\partial z^2}-\left(c_s^2+v_A^2\right)k_x^2-\frac{\partial^2}{\partial t^2}\right)v_x=k_x\left(c_s^2\frac{\partial}{\partial z}-g\right)v_z,
\label{wavx.eqn}
\end{equation}
\begin{equation}
\left(c_s^2\frac{\partial^2}{\partial z^2}-\gamma g\frac{\partial}{\partial z}-\frac{\partial^2}{\partial t^2}\right)v_z=-k_x\left(c_s^2\frac{\partial}{\partial z}-\left(\gamma-1\right)g\right)v_x,
\label{wavz.eqn}
\end{equation}
where $c_s^2=\gamma p_0/\rho_0$ and $v_A^2=B_0^2/\left(\mu\rho_0\right)$ are the squared sound and Alfv\'en speeds respectively.  These are in agreement with those discovered by \cite{Ferraro1958}.  

Examining these equations we can see that if the effects of gravity are negligible and the horizontal wavenumber is small, then the wave operators on the left-hand side are the same when $c_s=v_A$.  At this point there is a resonance which allows the mode conversion process to occur.  Note that when the horizontal wavenumber is zero, there is no mode conversion.  

We solve these equations numerically using the MacCormack finite difference scheme.  This is a type of Lax-Wendroff, two-step, predictor-corrector method.  We use backward differencing for the predictor steps and forward differencing for the corrector steps, thus ensuring that we have more accurate values on the upper boundary where we drive a sine wave in the vertical velocity $v_z$.  As previously mentioned, this is predominantly a slow wave as the fast wave is evanescent in the low-$\beta$ plasma.  

\begin{table}[!h]
\begin{center}
\begin{tabular}{l c c}
\hline
 & High $\beta$ & Low $\beta$ \\
\hline
Slow Wave & Speed$\sim v_A$ & Speed$\sim c_s$ \\
 & $\perp$ to {\bf B} & $\parallel$ to {\bf B} \\
\hline
Fast Wave & Speed$\sim c_s$ & Speed$\sim v_A$ \\
 & Approx.~isotropic & $\perp$ to {\bf B}\\
\hline
\end{tabular}
\caption{Characteristic properties of the slow and fast magnetoacoustic waves depending on their position in the solar atmosphere.}
\label{wavprop.tab}
\end{center}
\end{table}

As the wave propagates across the $\beta\approx1$ layer it is important to remember that the slow and fast waves have different properties above and below this region; Table~\ref{wavprop.tab} identifies these.  We are driving a slow wave in low-$\beta$ plasma, the transmitted component of this wave will demonstrate the same characteristics in the high-$\beta$ plasma.  From the table we see that the transmitted wave is thus the fast wave, and the converted part must then be the slow wave.

\section{Isothermal Atmosphere}
\label{iso.sec}

To begin with we investigate an isothermal atmosphere, so the sound speed will remain constant while the Alfv\'en speed varies with height.  

\subsection{Numerical Simulations}

\begin{figure}[t]
\centering
\includegraphics[angle=90,scale=0.56]{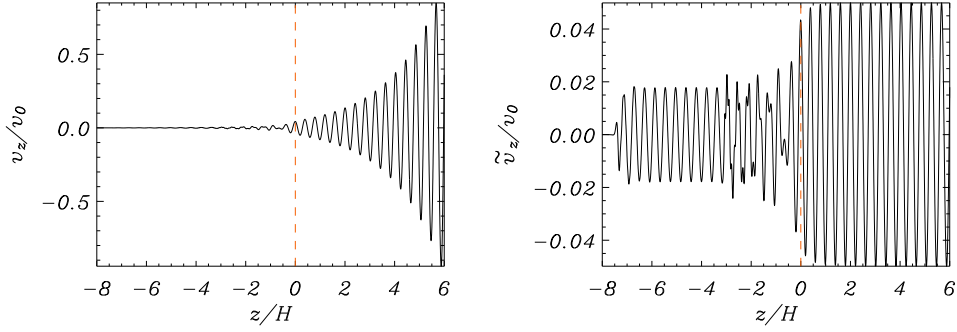}
\caption{{\it Left:} Vertical velocity, $v_z$, for $\omega=2\upi\sqrt{6}$ and $k_x=\upi$ at $t=13.5$ Alfv\'en times.  \break {\it Right:} Transformed vertical velocity, $\tilde{v}_z$, for the same parameters.  In both cases the dashed line indicates where $c_s=v_A$.}
\label{iso.fig}
\end{figure}

The left-hand plot in Figure~\ref{iso.fig} shows the vertical velocity from the numerical simulation.  This displays a strong exponential nature which is due to the gravitational stratification, and this masks what is occurring at the mode-conversion region (dashed line) which lies at $z=0$.  We may overcome this by making the transformation $v_z\rightarrow\tilde{v}_ze^{z/2}$ so that the exponential behaviour is removed and the incident wave has a constant amplitude.  Upon doing this we obtain the right-hand plot of Figure~\ref{iso.fig}.  There is now a clear change at the conversion region, with a definite decrease in amplitude and the presence of two \pagebreak wave modes.  The transmitted fast wave can be seen propagating out in front, while the converted slow wave is visible in the form of interference.  As both the incident and transmitted waves have a constant amplitude after this transformation we may calculate the proportion of transmission directly.  We shall utilise this later to test the accuracy of the analytical predictions.  

\subsection{Analytical Study}

The first analytical technique we use to examine mode conversion is the WKB method.  This involves expanding the horizontal and vertical velocity components, $v_x$ and $v_z$, in inverse powers of $\omega$ (the driving frequency).  This is valid under the assumption that $\omega$ is large, and will give solutions valid outside the mode-conversion region.  Substituting these expansions into the Wave Equations~(\ref{wavx.eqn}) and~(\ref{wavz.eqn}) we obtain the solutions
\begin{eqnarray*}
\textrm{Low} \beta\textrm{:}\qquad & \textrm{Incident Wave}\qquad & v_z\propto e^{z/2}\exp{\left(\frac{i\omega z}{c_s}\right)},\\
\textrm{High} \beta\textrm{:}\qquad & \textrm{Transmitted Wave}\qquad & v_z\propto Be^{z/2}\exp{\left(\frac{i\omega z}{c_s}\right)},\\
 & \textrm{Converted Wave}\qquad & v_z\propto -A\frac{ik_xc_s^2v_Ae^{z/4}}{\omega\left(v_A^2-c_s^2\right)}\exp{\left(-\frac{2i\omega}{v_A}\right)},
\end{eqnarray*}
where $A$ and $B$ are the conversion and transmission coefficients respectively.  Note that we have also extracted the exponential dependence seen in the simulations.  

These coefficients are the same as those found previously by \cite[Zhugzhda \& Dzhalilov]{Zhugzhda1982}, which were possible to determine because the exact analytical solution was known.  We use a method developed by \cite{Cairns1983} to find the coefficients.  This method is valid at the mode-conversion region, and has the advantage that coefficients may be found even when there is no exact solution.  We shall take advantage of this when we look at the non-isothermal atmosphere in Section~\ref{noniso.sec}.

From \cite{Cairns1983}, we can write the Wave Equations~(\ref{wavx.eqn}) and~(\ref{wavz.eqn}) in the general form
\begin{equation}
\frac{\textrm{d}\phi_1}{\textrm{d}\xi}-i\left(k_0-\frac{b}{a}\xi\right)\phi_1=i\lambda\phi_2,
\label{cairnsx.eqn}
\end{equation}
\begin{equation}
\frac{\textrm{d}\phi_2}{\textrm{d}\xi}-i\left(k_0-\frac{g}{f}\xi\right)\phi_2=i\lambda\phi_1,
\label{cairnsz.eqn}
\end{equation}
where $\phi_1$ and $\phi_2$ are the main variables, and the conversion and transmission coefficients are given by
\begin{equation}
A=\sqrt{1-\exp{\left(-\frac{2\upi\lambda^2af}{ag-bf}\right)}},\qquad B=\exp{\left(-\frac{\upi\lambda^2af}{ag-bf}\right)}.
\end{equation}
In order to manipulate Equations~(\ref{wavx.eqn}) and~(\ref{wavz.eqn}) into the correct format we take Fourier components in time, we must then impose the conditions that $k_x$ is small and $\omega$ is large, allowing us to neglect some terms.  Finally making the transformation $v_z=ic_sV_z/v_A$ and expanding $z=0+\xi$ about the mode-conversion region our equations may be written
\begin{equation}
\frac{\textrm{d}v_x}{\textrm{d}\xi}-i\left(\frac{\omega}{c_s}-\frac{\omega}{2Hc_s}\xi\right)v_x=\frac{ik_x}{2}V_z,
\end{equation}
\begin{equation}
\frac{\textrm{d}V_z}{\textrm{d}\xi}-\frac{i\omega}{c_s}V_z=\frac{ik_x}{2}v_x.
\end{equation}
By comparing these to Equations~(\ref{cairnsx.eqn}) and~(\ref{cairnsz.eqn}) we may simply write down the conversion and transmission coefficients
\begin{equation}
A=\sqrt{1-\exp{\left(-\frac{\upi k_x^2c_sH}{\omega}\right)}},\qquad B=\exp{\left(-\frac{\upi k_x^2c_sH}{2\omega}\right)}.
\label{conv.eqn}
\end{equation}

\begin{figure}[t]
\centering
\includegraphics[scale=0.4]{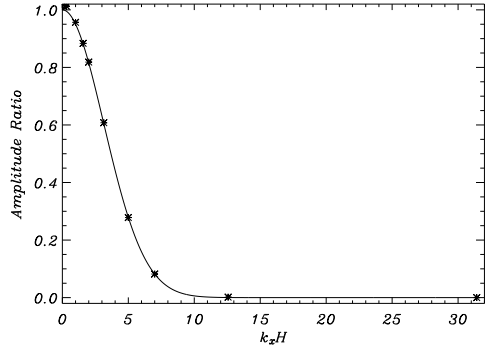}
\caption{Ratio of the slow wave amplitudes in high and low $\beta$, telling us how much of the incident wave will be transmitted into the high-$\beta$ plasma.  Here $\omega=4\upi\sqrt{6}$, the solid line is that predicted analytically, and the stars are the values calculated from the numerical data.}
\label{iamprat.fig}
\end{figure}

As the amount of transmission is easily calculated from the numerical simulations we may compare this directly to what is predicted analytically by Equation~(\ref{conv.eqn}).  We did this for various values of $k_x$ and the results are shown in Figure~\ref{iamprat.fig} where the stars indicate the transmission in the numerical simulations and the solid line is that predicted analytically.  The agreement between the numerical and analytical results is excellent, even when $k_x$ becomes large thus violating our assumptions.  This suggests that the analytical predictions are an excellent tool for quantifying mode conversion.  More information on the analytical techniques may be found in \cite{McDougall2007}.  

\section{Non-Isothermal Atmosphere}
\label{noniso.sec}

\begin{figure}[t]
\centering
\includegraphics[scale=0.4]{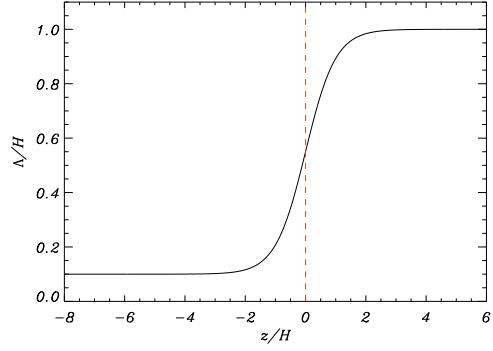}
\caption{A temperature profile of the form $\Lambda=a+b\tanh{z}$ is chosen to mimic the steep gradient of the transition region.  Here we have set $a=0.55$ and $b=0.45$, and the dashed line indicates where $c_s=v_A$.}
\label{temprof.fig}
\end{figure}

Exactly the same analysis may be carried out for a non-isothermal atmosphere, where there is no exact analytical solution.  We choose a $\tanh$ curve for the temperature profile, as shown in Figure~\ref{temprof.fig}, as this mimics the steep temperature gradient of the transition region.  The form of this profile is $\Lambda=a+b\tanh{\left(z/H\right)}$ and $c_s=v_A$ is chosen to lie in the centre of the steep gradient at $z=0$, as indicated by the dashed line.  We may use Wave Equations~(\ref{wavx.eqn}) and~(\ref{wavz.eqn}) here, but must remember that both the sound and Alfv\'en speeds will now vary with height.  

\subsection{Analytical Study}

\begin{figure}[t]
\centering
\includegraphics[angle=90,scale=0.56]{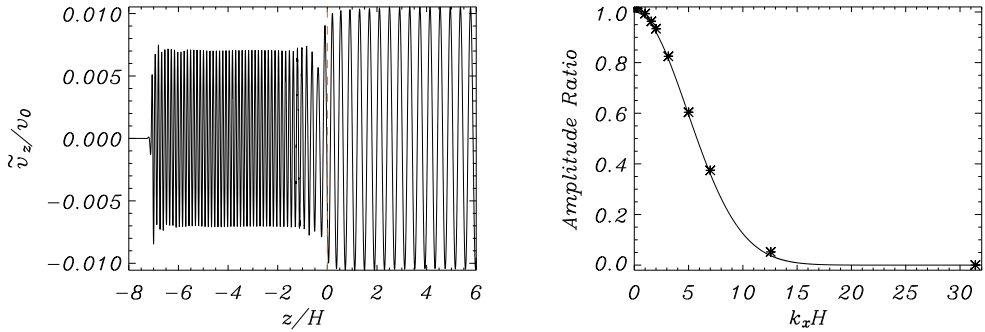}
\caption{\textit{Left:} Transformed vertical velocity, $\tilde{v}_z$, for $\omega=2\upi\sqrt{6}\sqrt{\gamma\beta_0/2}$ and $k_x=\upi$ at $t=7$ Alfv\'en times.  The dashed line indicates where $c_s=v_A$.  \break\textit{Right:} Ratio of the slow wave amplitudes in high and low $\beta$, telling us how much of the incident wave will be transmitted into the high-$\beta$ plasma.  Here $\omega=4\upi\sqrt{6}\sqrt{\gamma\beta_0/2}$, the solid line is that predicted analytically, and the stars are the values calculated from the numerical data.}
\label{niso.fig}
\end{figure}

Using a WKB analysis on Equations~(\ref{wavx.eqn}) and~(\ref{wavz.eqn}) the behaviour away from the mode-conversion region is given by
\begin{equation}
v_z\propto\frac{\Lambda^{1/4}}{p_0^{1/2}}\exp{\left(i\omega\sqrt{\frac{2}{\gamma\beta_0}}\int{\frac{1}{c_s}\textrm{d}z}\right)},
\end{equation}
where $\beta_0$ is the plasma beta at $z=0$.  This equation tells us that to obtain a constant amplitude in the incident and transmitted waves, as we had in the isothermal case, we must make the transformation $v_z\rightarrow \tilde{v}_z\Lambda^{1/4}/p_0^{1/2}$ (shown in the left-hand plot of Figure~\ref{niso.fig}).  Note that if we set $\Lambda=1$ we revert back to an isothermal atmosphere with the same equations as Section~\ref{iso.sec}.  

We may then follow \cite[Cairns \& Lashmore-Davies]{Cairns1983} to find the behaviour around the conversion region, giving exactly the same conversion and transmission coefficients as before.  So varying the temperature profile has no effect on the amount of transmission and conversion as a wave propagates through the mode-conversion region.  

As in the isothermal case, the amount of transmission may be calculated directly from the numerical results, so we compare this to the analytical prediction for various values of $k_x$.  The agreement between the two is again excellent, as illustrated in the right-hand plot of Figure~\ref{niso.fig}.

\section{Conclusions}
\label{conc.sec}

Using a combination of analytical techniques and numerical computations we have studied the phenomenon of mode conversion in a simple, one-dimensional model.  More specifically, we have studied the downward propagation of linear waves in a gravitationally stratified atmosphere permeated by a uniform, vertical magnetic field.  These waves propagate from low- to high-$\beta$ plasma, and the mode conversion occurs where the sound and Alfv\'en speeds are equal where $\beta\approx1$.  

In Section~\ref{iso.sec} we concentrated on an isothermal atmosphere using a combination of the WKB method, and a method developed by \cite[Cairns \& Lashmore-Davies]{cairns1983} to examine mode conversion both outside and around the conversion region.  This allowed conversion and transmission coefficients to be found.  It was then possible to compare the transmission predicted by the analytical method with the numerical data.  These showed excellent agreement, even as $k_x$ becomes large, violating the assumptions.  

This analysis was repeated in Section~\ref{noniso.sec} for a non-isothermal atmosphere.  We selected a $\tanh$ temperature profile to mimic the steep temperature gradient of the transition region.  Surprisingly the transmission and conversion coefficients found were identical to those for the isothermal atmosphere.  So the temperature profile has no effect on the amount of transmission and conversion as a wave undergoes mode conversion.  

Although the model set-up is not physically realistic, it has given us a great insight into the mode conversion process, and the techniques which may be utilised in its analysis.  We plan to use this knowledge to look at a more realistic model of a two-dimensional coronal null point, extending on work done by \cite{McLaughlin2006}.  In this situation the plasma $\beta$ is infinite at the null point, so as the wave propagates towards the null it will indeed be passing from low- to high-$\beta$ plasma, as we have examined here.  

\begin{acknowledgments}
Dee McDougall acknowledges financial assistance from the Carnegie Trust for the Universities of Scotland and a travel grant from the International Astronomical Union.  
\end{acknowledgments}

\end{document}